\documentclass[12pt,reqno]{amsart}
\usepackage{graphicx}
\usepackage{amscd}
\usepackage{amsmath}
\usepackage{epsfig}
\usepackage{amsfonts}
\usepackage{amssymb}

\providecommand{\U}[1]{\protect\rule{.1in}{.1in}}
\providecommand{\U}[1]{\protect\rule{.1in}{.1in}}
\allowdisplaybreaks

\theoremstyle{plain}

\numberwithin{equation}{section}

\input{tcilatex}

\begin{document}
\title[Damped Oscillators in Quantum Mechanics]{Models of Damped Oscillators
in Quantum Mechanics}
\author{Ricardo Cordero-Soto}
\address{Mathematical, Computational and Modeling Sciences Center, Arizona
State University, Tempe, AZ 85287--1804, U.S.A.}
\email{ricardojavier81@gmail.com}
\author{Erwin Suazo}
\address{School of Mathematical and Statistical Sciences, Mathematical,
Computational and Modeling Sciences Center, Arizona State University, Tempe,
AZ 85287--1804, U.S.A.}
\email{suazo@mathpost.la.asu.edu}
\author{Sergei K. Suslov}
\address{School of Mathematical and Statistical Sciences, Mathematical,
Computational and Modeling Sciences Center, Arizona State University, Tempe,
AZ 85287--1804, U.S.A.}
\email{sks@asu.edu}
\urladdr{http://hahn.la.asu.edu/\symbol{126}suslov/index.html}
\date{\today }

\begin{abstract}
We consider several models of the damped oscillators in nonrelativistic
quantum mechanics in a framework of a general approach to the dynamics of
the time-dependent Schr\"{o}dinger equation with variable quadratic
Hamiltonians. The Green functions are explicitly found in terms of
elementary functions and the corresponding gauge transformations are
discussed. The factorization technique is applied to the case of a shifted
harmonic oscillator. The time-evolution of the expectation values of the
energy related operators is determined for two models of the quantum damped
oscillators under consideration. The classical equations of motion for the
damped oscillations are derived for the corresponding expectation values of
the position operator.
\end{abstract}

\maketitle

\section{An Introduction}

We continue an investigation of the one-dimensional Schr\"{o}dinger
equations with variable quadratic Hamiltonians of the form%
\begin{equation}
i\frac{\partial \psi }{\partial t}=-a\left( t\right) \frac{\partial ^{2}\psi 
}{\partial x^{2}}+b\left( t\right) x^{2}\psi -i\left( c\left( t\right) x%
\frac{\partial \psi }{\partial x}+d\left( t\right) \psi \right) ,
\label{in1}
\end{equation}%
where $a\left( t\right) ,$ $b\left( t\right) ,$ $c\left( t\right) ,$ and $%
d\left( t\right) $ are real-valued functions of time $t$ only; see Refs.~%
\cite{Cor-Sot:Lop:Sua:Sus}, \cite{Cor-Sot:Sus}, \cite{Lan:Sus}, \cite%
{Lop:Sus}, \cite{Me:Co:Su}, \cite{Sua:Sus}, \cite{Suaz:Sus}, and \cite%
{Sua:Sus:Vega} for a general approach and currently known explicit
solutions. Here we discuss elementary cases related to the models of damped
oscillators. The corresponding Green functions, or Feynman's propagators,
can be found as follows \cite{Cor-Sot:Lop:Sua:Sus}, \cite{Suaz:Sus}:%
\begin{equation}
\psi =G\left( x,y,t\right) =\frac{1}{\sqrt{2\pi i\mu \left( t\right) }}\
e^{i\left( \alpha \left( t\right) x^{2}+\beta \left( t\right) xy+\gamma
\left( t\right) y^{2}\right) },  \label{in2}
\end{equation}%
where%
\begin{eqnarray}
&&\alpha \left( t\right) =\frac{1}{4a\left( t\right) }\frac{\mu ^{\prime
}\left( t\right) }{\mu \left( t\right) }-\frac{d\left( t\right) }{2a\left(
t\right) },  \label{in3} \\
&&\beta \left( t\right) =-\frac{h\left( t\right) }{\mu \left( t\right) }%
,\qquad h\left( t\right) =\exp \left( -\int_{0}^{t}\left( c\left( \tau
\right) -2d\left( \tau \right) \right) \ d\tau \right) ,  \label{in4} \\
&&\gamma \left( t\right) =\frac{a\left( t\right) h^{2}\left( t\right) }{\mu
\left( t\right) \mu ^{\prime }\left( t\right) }+\frac{d\left( 0\right) }{%
2a\left( 0\right) }-4\int_{0}^{t}\frac{a\left( \tau \right) \sigma \left(
\tau \right) h^{2}\left( \tau \right) }{\left( \mu ^{\prime }\left( \tau
\right) \right) ^{2}}\ d\tau ,  \label{in5}
\end{eqnarray}%
and the function $\mu \left( t\right) $ satisfies the\ characteristic
equation%
\begin{equation}
\mu ^{\prime \prime }-\tau \left( t\right) \mu ^{\prime }+4\sigma \left(
t\right) \mu =0  \label{in6}
\end{equation}%
with%
\begin{equation}
\tau \left( t\right) =\frac{a^{\prime }}{a}-2c+4d,\qquad \sigma \left(
t\right) =ab-cd+d^{2}+\frac{d}{2}\left( \frac{a^{\prime }}{a}-\frac{%
d^{\prime }}{d}\right)  \label{in7}
\end{equation}%
subject to the initial data%
\begin{equation}
\mu \left( 0\right) =0,\qquad \mu ^{\prime }\left( 0\right) =2a\left(
0\right) \neq 0.  \label{in8}
\end{equation}%
More details can be found in Refs.~\cite{Cor-Sot:Lop:Sua:Sus} and \cite%
{Suaz:Sus}. The corresponding Hamiltonian structure is discussed in Ref.~%
\cite{Cor-Sot:Sus}.\medskip

The simple harmonic oscillator is of interest in many advanced quantum
problems \cite{Fey:Hib}, \cite{La:Lif}, \cite{Merz}, and \cite{Schiff}. The
forced harmonic oscillator was originally considered by Richard Feynman in
his path integrals approach to the nonrelativistic quantum mechanics \cite%
{FeynmanPhD}, \cite{Feynman}, \cite{Feynman49a}, \cite{Feynman49b}, and \cite%
{Fey:Hib}; see also \cite{Lop:Sus}. Its special and limiting cases were
discussed by many authors; see Refs.~\cite{Beauregard}, \cite{Gottf:T-MY}, 
\cite{Holstein}, \cite{Maslov:Fedoriuk}, \cite{Merz}, \cite{Thomber:Taylor}
for the simple harmonic oscillator and Refs.~\cite{Arrighini:Durante}, \cite%
{Brown:Zhang}, \cite{Holstein97}, \cite{Nardone}, \cite{Robinett} for the
particle in a constant external field and references therein.\medskip

The damped oscillations have been analyzed to a great extent in classical
mechanics; see, for example, Refs.~\cite{BatemanPDE} and \cite{Lan:Lif}. In
the present paper we consider the time-dependent Schr\"{o}dinger equation%
\begin{equation}
i\frac{\partial \psi }{\partial t}=H\psi  \label{in9}
\end{equation}%
with the following nonself-adjoint Hamiltonians%
\begin{equation}
H=\frac{\omega _{0}}{2}\left( p^{2}+x^{2}\right) -\lambda px  \label{in10}
\end{equation}%
and%
\begin{equation}
H=\frac{\omega _{0}}{2}\left( p^{2}+x^{2}\right) -\lambda xp,  \label{in11}
\end{equation}%
where $p=-i\partial /\partial x,$ as quantum analogs of the damped
oscillator. A related self-adjoint Hamiltonian%
\begin{equation}
H=\frac{\omega _{0}}{2}\left( p^{2}+x^{2}\right) -\frac{\lambda }{2}\left(
px+xp\right)  \label{in12}
\end{equation}%
is also analyzed. Although discussion of a quantum damped oscillator is
usually missing in the standard classical textbooks \cite{La:Lif}, \cite%
{Merz}, and \cite{Schiff} among others, we believe that the models presented
here have a significant value from the pedagogical and mathematical points
of view. For instance, one of these models was crucial for our understanding
of a \textquotedblleft hidden\textquotedblright\ symmetry of the quadratic
propagators in Ref.~\cite{Cor-Sot:Sus}. Moreover, our models show that
fundamentals of quantum mechanics, such as evolution of the expectation
values of operators and Ehrenfest's theorem, can be extended to the case of
nonself-adjoint Hamiltonians. This provides, in our opinion, a somewhat
better understanding of the mathematical foundations of quantum mechanics
and can be used in the classroom.\medskip

The paper is organized as follows. In section~2 we derive the propagators
for the models of the damped oscillator (\ref{in10}) and (\ref{in11})
following the method of Ref.~\cite{Cor-Sot:Lop:Sua:Sus}. The corresponding
gauge transformations are discussed in section~3. The next section is
concerned with the separation of the variables for related model of a
\textquotedblleft shifted\textquotedblright\ linear harmonic oscillator (\ref%
{in12}). The factorization technique is applied to this oscillator in
section~5. The time evolution of the expectation values of the energy
related operators is determined for these quantum damped oscillators in
section~6. The classical equations for the damped oscillations are derived
for the expectation values of the position operator in the next section. One
more model of the damped oscillator with a variable quadratic Hamiltonian is
introduced in section~8. The last section contains some remarks on the
momentum representation.

\section{The First Two Models}

For the time-dependent Schr\"{o}dinger equation:%
\begin{equation}
i\frac{\partial \psi }{\partial t}=\frac{\omega _{0}}{2}\left( -\frac{%
\partial ^{2}\psi }{\partial x^{2}}+x^{2}\psi \right) +i\lambda \left( x%
\frac{\partial \psi }{\partial x}+\psi \right)  \label{fm1}
\end{equation}%
with $a=b=\omega _{0}/2$ and $c=d=-\lambda ,$ the characteristic equation
takes the form of the classical equation of motion for the damped oscillator 
\cite{BatemanPDE}, \cite{Lan:Lif}: 
\begin{equation}
\mu ^{\prime \prime }+2\lambda \mu ^{\prime }+\omega _{0}^{2}\mu =0,
\label{fm2}
\end{equation}%
whose suitable solution is as follows%
\begin{equation}
\mu =\frac{\omega _{0}}{\omega }e^{-\lambda t}\sin \omega t,\qquad \omega =%
\sqrt{\omega _{0}^{2}-\lambda ^{2}}>0.  \label{fm3}
\end{equation}%
The corresponding propagator is given by%
\begin{eqnarray}
G\left( x,y,t\right) &=&\sqrt{\frac{\omega e^{\lambda t}}{2\pi i\omega
_{0}\sin \omega t}}\exp \left( \frac{i\omega }{2\omega _{0}\sin \omega t}%
\left( \left( x^{2}+y^{2}\right) \cos \omega t-2xy\right) \right)  \notag \\
&&\times \exp \left( \frac{i\lambda }{2\omega _{0}}\left( x^{2}-y^{2}\right)
\right) .  \label{fm4}
\end{eqnarray}%
Indeed, directly from (\ref{in3})--(\ref{in4}):%
\begin{equation}
\alpha \left( t\right) =\frac{\omega \cos \omega t+\lambda \sin \omega t}{%
2\omega _{0}\sin \omega t},\qquad \beta \left( t\right) =-\frac{\omega }{%
\omega _{0}\sin \omega t}.  \label{fm5}
\end{equation}%
The integral in (\ref{in5}) can be evaluated with the help of a familiar
antiderivative%
\begin{equation}
\int \frac{dt}{\left( A\cos t+B\sin t\right) ^{2}}=\frac{\sin t}{A\left(
A\cos t+B\sin t\right) }+C.  \label{fm6}
\end{equation}%
It gives%
\begin{equation}
\gamma \left( t\right) =\frac{\omega \cos \omega t-\lambda \sin \omega t}{%
2\omega _{0}\sin \omega t}  \label{fm7}
\end{equation}%
with the help of the following identity%
\begin{equation}
\omega ^{2}-\omega _{0}^{2}\sin ^{2}\omega t=\omega ^{2}\cos ^{2}\omega
t-\lambda ^{2}\sin ^{2}\omega t  \label{fm7a}
\end{equation}%
and the propagator (\ref{fm4}) is verified. A \textquotedblleft
hidden\textquotedblright\ symmetry of this propagator is discussed in Ref.~%
\cite{Cor-Sot:Sus}.\medskip

The time-evolution of the squared norm of the wave function is given by%
\begin{equation}
\left\Vert \psi \left( x,t\right) \right\Vert ^{2}=\int_{-\infty }^{\infty
}\left\vert \psi \left( x,t\right) \right\vert ^{2}\ dx=e^{\lambda
t}\left\Vert \psi \left( x,0\right) \right\Vert ^{2}.  \label{fm8}
\end{equation}%
It is derived in section~6 among other things. We have discussed here the
case $\omega _{0}^{2}>\lambda ^{2}.$ Two more cases, when $\omega
_{0}^{2}=\lambda ^{2}$ and $\omega _{0}^{2}<\lambda ^{2},$ are similar and
the details are left to the reader.\medskip

In a similar fashion, the time-dependent Schr\"{o}dinger equation of the form%
\begin{equation}
i\frac{\partial \psi }{\partial t}=\frac{\omega _{0}}{2}\left( -\frac{%
\partial ^{2}\psi }{\partial x^{2}}+x^{2}\psi \right) +i\lambda x\frac{%
\partial \psi }{\partial x}  \label{fm9}
\end{equation}%
with $a=b=\omega _{0}/2$ and $c=-\lambda ,$ $d=0,$ has the characteristic
equation%
\begin{equation}
\mu ^{\prime \prime }-2\lambda \mu ^{\prime }+\omega _{0}^{2}\mu =0
\label{fm10}
\end{equation}%
with the solution%
\begin{equation}
\mu =\frac{\omega _{0}}{\omega }e^{\lambda t}\sin \omega t,\qquad \omega =%
\sqrt{\omega _{0}^{2}-\lambda ^{2}}>0.  \label{fm11}
\end{equation}%
The corresponding propagator is given by%
\begin{eqnarray}
G\left( x,y,t\right) &=&\sqrt{\frac{\omega e^{-\lambda t}}{2\pi i\omega
_{0}\sin \omega t}}\exp \left( \frac{i\omega }{2\omega _{0}\sin \omega t}%
\left( \left( x^{2}+y^{2}\right) \cos \omega t-2xy\right) \right)  \notag \\
&&\times \exp \left( \frac{i\lambda }{2\omega _{0}}\left( x^{2}-y^{2}\right)
\right)  \label{fm12}
\end{eqnarray}%
and the evolution of the squared norm is as follows%
\begin{equation}
\left\Vert \psi \left( x,t\right) \right\Vert ^{2}=e^{-\lambda t}\left\Vert
\psi \left( x,0\right) \right\Vert ^{2}.  \label{fm12a}
\end{equation}

The solution of the Cauchy initial value problem%
\begin{equation}
i\frac{\partial \psi }{\partial t}=H\psi ,\qquad \psi \left( x,0\right)
=\chi \left( x\right)  \label{ivp}
\end{equation}%
for our models (\ref{fm1}) and (\ref{fm9}) is given by the superposition
principle in an integral form%
\begin{equation}
\psi \left( x,t\right) =\int_{-\infty }^{\infty }G\left( x,y,t\right) \ \chi
\left( y\right) \ dy  \label{supp}
\end{equation}%
for a suitable initial function $\chi $ on $\boldsymbol{R};$ a rigorous
proof is given in Ref.~\cite{Suaz:Sus}.

\section{The Gauge Transformations}

The time-dependent Schr\"{o}dinger equation%
\begin{equation}
i\frac{\partial \psi }{\partial t}=\left( \frac{\omega _{0}}{2}\left(
p-A\right) ^{2}+U+\left( p-A\right) V+W\left( p-A\right) \right) \psi ,
\label{g1}
\end{equation}%
where $p=i^{-1}\partial /\partial x$ is the linear momentum operator and $%
A=A\left( x,t\right) ,$ $U=U\left( x,t\right) ,$ $V=V\left( x,t\right) ,$ $%
W=W\left( x,t\right) $ are real-valued functions, with the help of the gauge
transformation%
\begin{equation}
\psi =e^{-if\left( x,t\right) }\psi ^{\prime }  \label{g2}
\end{equation}%
can be transformed into a similar form%
\begin{equation}
i\frac{\partial \psi ^{\prime }}{\partial t}=\left( \frac{\omega _{0}}{2}%
\left( p-A^{\prime }\right) ^{2}+U^{\prime }+\left( p-A^{\prime }\right)
V^{\prime }+W^{\prime }\left( p-A^{\prime }\right) \right) \psi ^{\prime }
\label{g3}
\end{equation}%
with the new vector and scalar potentials given by%
\begin{equation}
A^{\prime }=A+\frac{\partial f}{\partial x},\qquad U^{\prime }=U-\frac{%
\partial f}{\partial t},\qquad V^{\prime }=V,\qquad W^{\prime }=W.
\label{g4}
\end{equation}%
Here we consider the one-dimensional case only and may think of $f$ as being
an arbitrary complex-valued differentiable function. Also, the Hamiltonian
in the right hand side of equation (\ref{g1}) is not assumed to be
self-adjoint. See Refs.~\cite{La:Lif} and \cite{Merz} for discussion of the
traditional case, when $V=W\equiv 0.$\medskip

An interesting special case of the gauge transformation related to this
paper is given by%
\begin{eqnarray}
&&A=0,\qquad U=\frac{\omega _{0}}{2}x^{2},\qquad V=-\lambda x,\qquad
W=0,\qquad f=\frac{i\lambda t}{2},  \label{g5} \\
&&A^{\prime }=0,\qquad U^{\prime }=\frac{\omega _{0}}{2}x^{2}-\frac{i\lambda 
}{2},\qquad V^{\prime }=-\lambda x,\qquad W^{\prime }=0,  \label{g6}
\end{eqnarray}%
when the new Hamiltonian is%
\begin{eqnarray}
H^{\prime } &=&\frac{\omega _{0}}{2}\left( p-A^{\prime }\right)
^{2}+U^{\prime }+pV^{\prime }  \label{g7} \\
&=&\frac{\omega _{0}}{2}\left( -\frac{\partial ^{2}}{\partial x^{2}}%
+x^{2}\right) +i\frac{\lambda }{2}\left( 2x\frac{\partial }{\partial x}%
+1\right) ,  \notag
\end{eqnarray}%
and equation (\ref{fm1}) takes the form%
\begin{equation}
i\frac{\partial \psi }{\partial t}=\frac{\omega _{0}}{2}\left( -\frac{%
\partial ^{2}\psi }{\partial x^{2}}+x^{2}\psi \right) +i\frac{\lambda }{2}%
\left( 2x\frac{\partial \psi }{\partial x}+\psi \right) .  \label{g8}
\end{equation}

The corresponding Green function is given by%
\begin{eqnarray}
G\left( x,y,t\right) &=&\sqrt{\frac{\omega }{2\pi i\omega _{0}\sin \omega t}}%
\exp \left( \frac{i\omega }{2\omega _{0}\sin \omega t}\left( \left(
x^{2}+y^{2}\right) \cos \omega t-2xy\right) \right)  \notag \\
&&\times \exp \left( \frac{i\lambda }{2\omega _{0}}\left( x^{2}-y^{2}\right)
\right) ,\qquad \omega =\sqrt{\omega _{0}^{2}-\lambda ^{2}}>0  \label{g9}
\end{eqnarray}%
and the norm of the wave function is conserved with time. This can be
established once again directly from our equations (\ref{in2})--(\ref{in8}).
We leave the details to the reader. A traditional method of separation of
the variables and using the Mehler formula for Hermite polynomials is
discussed in the next section. The factorization technique is applied to
this Hamiltonian in section~5.\medskip

Equation (\ref{g8}), in turn, admits another local gauge transformation:%
\begin{eqnarray}
&&A=0,\qquad U=\frac{\omega _{0}}{2}x^{2},\qquad V=W=-\frac{\lambda x}{2}%
,\qquad f=-\frac{\lambda x^{2}}{2\omega _{0}},  \label{g10} \\
&&A^{\prime }=-\frac{\lambda x}{\omega _{0}},\qquad U^{\prime }=\frac{\omega
_{0}}{2}x^{2},\qquad V^{\prime }=W^{\prime }=-\frac{\lambda x}{2}
\label{g11}
\end{eqnarray}%
and the Hamiltonian becomes%
\begin{eqnarray}
H^{\prime } &=&\frac{\omega _{0}}{2}\left( p-A^{\prime }\right)
^{2}+U^{\prime }+\left( p-A^{\prime }\right) V^{\prime }+W^{\prime }\left(
p-A^{\prime }\right)  \notag \\
&=&\frac{\omega _{0}}{2}\left( p+\frac{\lambda x}{\omega _{0}}\right) ^{2}+%
\frac{\omega _{0}}{2}x^{2}  \notag \\
&&+\left( p+\frac{\lambda x}{\omega _{0}}\right) \left( -\frac{\lambda x}{%
\omega _{0}}\right) +\left( -\frac{\lambda x}{\omega _{0}}\right) \left( p+%
\frac{\lambda x}{\omega _{0}}\right)  \notag \\
&=&\frac{\omega _{0}}{2}p^{2}+\frac{\omega _{0}^{2}-\lambda ^{2}}{2\omega
_{0}}x^{2}.  \label{g12}
\end{eqnarray}%
As a result, equation (\ref{g8}) takes the form of equation for the harmonic
oscillator: 
\begin{equation}
i\frac{\partial \psi }{\partial t}=\frac{\omega _{0}}{2}\left( -\frac{%
\partial ^{2}\psi }{\partial x^{2}}+\frac{\omega ^{2}}{\omega _{0}^{2}}%
x^{2}\psi \right) ,\qquad \omega ^{2}=\omega _{0}^{2}-\lambda ^{2}>0
\label{g12a}
\end{equation}%
and can be solved, once again, by the traditional method of separation of
the variables or by the factorization technique.

\section{Separation of Variables for a Shifted Harmonic Oscillator}

We shall refer to the case (\ref{g8}) as one of a shifted linear harmonic
oscillator. The Ansatz%
\begin{equation}
\psi \left( x,t\right) =e^{-iEt}\varphi \left( x\right)  \label{m1}
\end{equation}%
in the time-dependent Schr\"{o}dinger equation results in the stationary Schr%
\"{o}dinger equation%
\begin{equation}
H\varphi =E\varphi  \label{m2}
\end{equation}%
with the Hamiltonian (\ref{g7}). The last equation, namely,%
\begin{equation}
-\varphi ^{\prime \prime }+x^{2}\varphi +\frac{i\lambda }{\omega _{0}}\left(
2x\varphi ^{\prime }+\varphi \right) =\frac{2E}{\omega _{0}}\varphi ,
\label{m3}
\end{equation}%
with the help of the substitution%
\begin{equation}
\varphi =\exp \left( \frac{i\lambda x^{2}}{2\omega _{0}}\right) u\left(
x\right)  \label{m4}
\end{equation}%
is reduced to the following equation%
\begin{equation}
-u^{\prime \prime }+\frac{\omega ^{2}}{\omega _{0}^{2}}x^{2}u=\frac{2E}{%
\omega _{0}}u.  \label{m5}
\end{equation}%
The change of the variable%
\begin{equation}
u\left( x\right) =v\left( \xi \right) ,\qquad x=\xi \sqrt{\frac{\omega _{0}}{%
\omega }}  \label{m6}
\end{equation}%
gives us the stationary Schr\"{o}dinger equation for the simple harmonic
oscillator \cite{La:Lif}, \cite{Merz}, \cite{Ni:Uv}, \cite{Schiff}: 
\begin{equation}
v^{\prime \prime }+\left( 2\varepsilon -\xi ^{2}\right) v=0  \label{m7}
\end{equation}%
with $\varepsilon =E/\omega ,$ whose eigenfunctions are given in terms of
the Hermite polynomials as follows%
\begin{equation}
v_{n}=C_{n}e^{-\xi ^{2}/2}H_{n}\left( \xi \right) ,  \label{m8}
\end{equation}%
and the corresponding eigenvalues are%
\begin{equation}
\varepsilon _{n}=n+\frac{1}{2},\qquad E_{n}=\omega \left( n+\frac{1}{2}%
\right) \qquad \left( n=0,1,2,...\right) .  \label{m9}
\end{equation}%
Thus the normalized wave functions of our shifted oscillator (\ref{g8}) are
given by 
\begin{equation}
\psi _{n}\left( x,t\right) =e^{-i\omega \left( n+1/2\right) t}\varphi
_{n}\left( x\right) ,  \label{m10}
\end{equation}%
where%
\begin{equation}
\varphi _{n}\left( x\right) =C_{n}\exp \left( \frac{i\lambda x^{2}}{2\omega
_{0}}\right) e^{-\xi ^{2}/2}H_{n}\left( \xi \right) ,\qquad \xi =x\sqrt{%
\frac{\omega }{\omega _{0}}}  \label{m10a}
\end{equation}%
and%
\begin{equation}
\left\vert C_{n}\right\vert ^{2}=\sqrt{\frac{\omega }{\omega _{0}}}\frac{1}{%
\sqrt{\pi }2^{n}n!}  \label{m11}
\end{equation}%
in view of the orthogonality relation%
\begin{equation}
\int_{-\infty }^{\infty }\varphi _{n}^{\ast }\left( x\right) \varphi
_{m}\left( x\right) \ dx=\delta _{nm}.  \label{m12}
\end{equation}%
We use the star for complex conjugate.\medskip

Solution of the initial value problem (\ref{ivp}) can be found by the
superposition principle in the form%
\begin{equation}
\psi \left( x,t\right) =\sum_{n=0}^{\infty }c_{n}\ \psi _{n}\left(
x,t\right) ,  \label{m13}
\end{equation}%
where%
\begin{equation}
\psi \left( x,0\right) =\chi \left( x\right) =\sum_{n=0}^{\infty }c_{n}\
\varphi _{n}\left( x\right)  \label{m14}
\end{equation}%
and%
\begin{equation}
c_{n}=\int_{-\infty }^{\infty }\varphi _{n}^{\ast }\left( y\right) \chi
\left( y\right) \ dy  \label{m15}
\end{equation}%
in view of the orthogonality property (\ref{m12}). Substituting (\ref{m15})
into (\ref{m13}) and changing the order of the summation and integration,
one gets%
\begin{equation}
\psi \left( x,t\right) =\int_{-\infty }^{\infty }G\left( x,y,t\right) \chi
\left( y\right) \ dy,  \label{m16}
\end{equation}%
where the Green function is given as the eigenfunction expansion:%
\begin{equation}
G\left( x,y,t\right) =\sum_{n=0}^{\infty }e^{-i\omega \left( n+1/2\right)
t}\varphi _{n}\left( x\right) \varphi _{n}^{\ast }\left( y\right) .
\label{m17}
\end{equation}%
This infinite series is summable with the help of the Poisson kernel for the
Hermite polynomials (Mehler's formula) \cite{Rain}:%
\begin{equation}
\sum_{n=0}^{\infty }\frac{H_{n}\left( x\right) H_{n}\left( y\right) }{2^{n}n!%
}r^{n}=\frac{1}{\sqrt{1-r^{2}}}\exp \left( \frac{2xyr-\left(
x^{2}+y^{2}\right) r^{2}}{1-r^{2}}\right) ,\quad \left\vert r\right\vert <1.
\label{m18}
\end{equation}%
The result is given, of course, by equation (\ref{g9}).

\section{The Factorization Method for Shifted Harmonic Oscillator}

It is worth applying the well-known factorization technique (see, for
example, \cite{As:Su1}, \cite{As:Su2}, \cite{At:Su}, \cite{Doung} and \cite%
{Merz}) to the Hamiltonian (\ref{g7}). The corresponding ladder operators
can be found in the forms%
\begin{eqnarray}
a &=&\left( \alpha +i\beta \right) x+\gamma \frac{\partial }{\partial x},
\label{f1} \\
a^{\dagger } &=&\left( \alpha -i\beta \right) x-\gamma \frac{\partial }{%
\partial x},  \label{f2}
\end{eqnarray}%
where $\alpha ,$ $\beta $ and $\gamma $ are real numbers to be determined as
follows. One gets%
\begin{eqnarray}
aa^{\dagger }\psi &=&\left( \alpha ^{2}+\beta ^{2}\right) x^{2}\psi +\left(
\alpha -i\beta \right) \gamma \psi -2i\beta \gamma x\frac{\partial \psi }{%
\partial x}-\gamma ^{2}\frac{\partial ^{2}\psi }{\partial x^{2}},  \label{f3}
\\
a^{\dagger }a\psi &=&\left( \alpha ^{2}+\beta ^{2}\right) x^{2}\psi -\left(
\alpha +i\beta \right) \gamma \psi -2i\beta \gamma x\frac{\partial \psi }{%
\partial x}-\gamma ^{2}\frac{\partial ^{2}\psi }{\partial x^{2}},  \label{f4}
\end{eqnarray}%
whence%
\begin{equation}
\left( aa^{\dagger }-a^{\dagger }a\right) \psi =2\alpha \gamma \psi
\label{f5}
\end{equation}%
and%
\begin{equation}
\frac{1}{2}\left( aa^{\dagger }+a^{\dagger }a\right) \psi =-\gamma ^{2}\frac{%
\partial ^{2}\psi }{\partial x^{2}}+\left( \alpha ^{2}+\beta ^{2}\right)
x^{2}\psi -i\beta \gamma \left( 2x\frac{\partial \psi }{\partial x}+\psi
\right) .  \label{f6}
\end{equation}%
The canonical commutation relation occurs and the Hamiltonian (\ref{g7})
takes the standard form:%
\begin{equation}
H=\frac{\omega }{2}\left( aa^{\dagger }+a^{\dagger }a\right) ,  \label{f7}
\end{equation}%
if%
\begin{equation}
2\alpha \gamma =1,\qquad \omega \left( \alpha ^{2}+\beta ^{2}\right) =\omega
\gamma ^{2}=\frac{1}{2}\omega _{0},\qquad \omega \beta \gamma =-\frac{1}{2}%
\lambda .  \label{f8}
\end{equation}%
The relation $\omega _{0}^{2}=\omega ^{2}+\lambda ^{2},$ which defines the
new oscillator frequency, holds. As a result, the explicit form of the
annihilation and creation operators is given by%
\begin{eqnarray}
\sqrt{2}a &=&\left( \sqrt{\frac{\omega }{\omega _{0}}}-\frac{i\lambda }{%
\sqrt{\omega _{0}\omega }}\right) x+\sqrt{\frac{\omega _{0}}{\omega }}\frac{%
\partial }{\partial x},  \label{f9} \\
\sqrt{2}a^{\dagger } &=&\left( \sqrt{\frac{\omega }{\omega _{0}}}+\frac{%
i\lambda }{\sqrt{\omega _{0}\omega }}\right) x-\sqrt{\frac{\omega _{0}}{%
\omega }}\frac{\partial }{\partial x}.  \label{f10}
\end{eqnarray}%
The special case $\lambda =0$ and $\omega =\omega _{0}$ gives a traditional
form of these operators.\medskip\ 

The oscillator spectrum (\ref{m9}) and the corresponding stationary wave
functions (\ref{m10a}) can be obtain now in a standard way by using the
Heisenberg--Weyl algebra of the rasing and lowering operators. In addition,
the $n$-dimensional oscillator wave functions form a basis of the
irreducible unitary representation of the Lie algebra of the noncompact
group $SU\left( 1,1\right) $ corresponding to the discrete positive series $%
\mathcal{D}_{+}^{j};$ see \cite{Me:Co:Su}, \cite{Ni:Su:Uv} and \cite%
{Smir:Shit}.\smallskip\ Our operators (\ref{f9})--(\ref{f10}) allow us to
extend these group-theoretical properties for the case of the shifted
oscillators. We leave the details to the reader.

\section{Dynamics of Energy Related Expectation Values}

The expectation value of an operator $A$ in quantum mechanics is given by
the formula%
\begin{equation}
\left\langle A\right\rangle =\int_{-\infty }^{\infty }\psi ^{\ast }\left(
x,t\right) \ A\left( t\right) \psi \left( x,t\right) \ dx,  \label{ex1}
\end{equation}%
where the wave function satisfies the time-dependent Schr\"{o}dinger equation%
\begin{equation}
i\frac{\partial \psi }{\partial t}=H\psi .  \label{ex2}
\end{equation}%
The time derivative of this expectation value can be written as%
\begin{equation}
i\frac{d}{dt}\left\langle A\right\rangle =i\left\langle \frac{\partial A}{%
\partial t}\right\rangle +\left\langle AH-H^{\dagger }A\right\rangle ,
\label{ex3}
\end{equation}%
where $H^{\dagger }$ is the Hermitian adjoint of the Hamiltonian operator $%
H. $ Our formula is a simple extension of the well-known expression \cite%
{La:Lif}, \cite{Merz}, \cite{Schiff} to the case of a nonself-adjoint
Hamiltonian.\medskip

We apply formula (\ref{ex3}) to the Hamiltonian%
\begin{equation}
H=\frac{\omega _{0}}{2}\left( p^{2}+x^{2}\right) -\lambda px,\qquad p=-i%
\frac{\partial }{\partial x}  \label{ex4}
\end{equation}%
in equation (\ref{fm1}). A few examples will follow. In the case of the
identity operator $A=1,$ one gets%
\begin{equation}
AH-H^{\dagger }A=\lambda \left( xp-px\right) =i\lambda  \label{ex5}
\end{equation}%
by the Heisenberg commutation relation%
\begin{equation}
\left[ x,p\right] =xp-px=i.  \label{ex6}
\end{equation}%
As a result,%
\begin{equation}
\frac{d}{dt}\left\Vert \psi \right\Vert ^{2}=\lambda \left\Vert \psi
\right\Vert ^{2},  \label{ex7}
\end{equation}%
and time-evolution of the squared norm of the wave function for our model of
the damped quantum oscillator is given by equation (\ref{fm8}).\medskip

In a similar fashion, if $A=H,$ then%
\begin{equation}
H^{2}-H^{\dagger }H=\left( H-H^{\dagger }\right) H=i\lambda H,  \label{ex7a}
\end{equation}%
and%
\begin{equation}
\frac{d}{dt}\left\langle H\right\rangle =\lambda \left\langle H\right\rangle
,\qquad \left\langle H\right\rangle =\left\langle H\right\rangle
_{0}e^{\lambda t}.  \label{ex7b}
\end{equation}%
Moreover,%
\begin{equation}
\frac{d}{dt}\left\langle H^{n}\right\rangle =\lambda \left\langle
H^{n}\right\rangle ,\qquad \left\langle H^{n}\right\rangle =\left\langle
H^{n}\right\rangle _{0}e^{\lambda t}\qquad \left( n=0,1,2,...\right) ,
\label{ex7c}
\end{equation}%
which unifies the both of the previous cases.

Now we choose $A=p^{2},$ $A=x^{2}$ and $A=px+xp,$ respectively, in order to
obtain the following system:%
\begin{eqnarray}
&&\frac{d}{dt}\left\langle p^{2}\right\rangle =3\lambda \left\langle
p^{2}\right\rangle -\omega _{0}\left\langle px+xp\right\rangle ,  \notag \\
&&\frac{d}{dt}\left\langle x^{2}\right\rangle =-\lambda \left\langle
x^{2}\right\rangle +\omega _{0}\left\langle px+xp\right\rangle ,
\label{ex12} \\
&&\frac{d}{dt}\left\langle px+xp\right\rangle =2\omega _{0}\left(
\left\langle p^{2}\right\rangle -\left\langle x^{2}\right\rangle \right)
+\lambda \left\langle px+xp\right\rangle .  \notag
\end{eqnarray}%
Indeed,%
\begin{eqnarray}
p^{2}H-H^{\dagger }p^{2} &=&\frac{\omega _{0}}{2}\left[ p^{2},x^{2}\right]
+\lambda \left[ x,p^{3}\right]  \label{com1} \\
&=&3i\lambda p^{2}-i\omega _{0}\left( px+xp\right) ,  \notag
\end{eqnarray}%
\begin{eqnarray}
x^{2}H-H^{\dagger }x^{2} &=&\frac{\omega _{0}}{2}\left[ x^{2},p^{2}\right]
-\lambda x\left[ x,p\right] x  \label{com2} \\
&=&i\omega _{0}\left( px+xp\right) -i\lambda x^{2},  \notag
\end{eqnarray}%
and%
\begin{eqnarray}
&&\left( px+xp\right) H-H^{\dagger }\left( px+xp\right)  \label{com3} \\
&&\quad =\frac{\omega _{0}}{2}\left( \left[ p,x^{3}\right] +\left[ x,p^{3}%
\right] \right)  \notag \\
&&\qquad +\frac{\omega _{0}}{2}\left( p\left[ x,p\right] p-x\left[ x,p\right]
x\right)  \notag \\
&&\qquad \quad +\lambda \left( \left( xp\right) ^{2}-\left( px\right)
^{2}\right)  \notag \\
&&\quad =2i\omega _{0}\left( p^{2}-x^{2}\right) +i\lambda \left(
px+xp\right) ,  \notag
\end{eqnarray}%
which results in (\ref{ex12}).\medskip

The system can be solved explicitly, thus providing the complete dynamics of
these expectation values. The eigenvalues are given by $r_{0}=\lambda ,$ $%
r_{\pm }=\lambda \pm 2i\omega $ and the corresponding linearly independent
eigenvectors are%
\begin{equation}
\boldsymbol{x}_{0}=\left( 
\begin{array}{c}
\omega _{0} \\ 
\omega _{0} \\ 
2\lambda%
\end{array}%
\right) ,\qquad \boldsymbol{x}_{\pm }=\left( 
\begin{array}{c}
\left( \lambda \pm i\omega \right) ^{2} \\ 
\omega _{0}^{2} \\ 
2\omega _{0}\left( \lambda \pm i\omega \right)%
\end{array}%
\right)  \label{ex12a}
\end{equation}%
with the determinant%
\begin{equation}
\left\vert 
\begin{array}{ccc}
\omega _{0} & \left( \lambda +i\omega \right) ^{2} & \left( \lambda -i\omega
\right) ^{2} \\ 
\omega _{0} & \omega _{0}^{2} & \omega _{0}^{2} \\ 
2\lambda & 2\omega _{0}\left( \lambda +i\omega \right) & 2\omega _{0}\left(
\lambda -i\omega \right)%
\end{array}%
\right\vert =-8i\omega _{0}^{2}\omega ^{3}\neq 0.  \label{ex12b}
\end{equation}%
The general solution of the system (\ref{ex12}) can be obtain in a complex
form as follows%
\begin{eqnarray}
&&\left( 
\begin{array}{c}
\left\langle p^{2}\right\rangle \\ 
\left\langle x^{2}\right\rangle \\ 
\left\langle px+xp\right\rangle%
\end{array}%
\right) =C_{0}e^{\lambda t}\left( 
\begin{array}{c}
\omega _{0} \\ 
\omega _{0} \\ 
2\lambda%
\end{array}%
\right)  \label{ex12c} \\
&&\quad \quad +C_{+}e^{\left( \lambda +2i\omega \right) t}\left( 
\begin{array}{c}
\left( \lambda +i\omega \right) ^{2} \\ 
\omega _{0}^{2} \\ 
2\omega _{0}\left( \lambda +i\omega \right)%
\end{array}%
\right) +C_{-}e^{\left( \lambda -2i\omega \right) t}\left( 
\begin{array}{c}
\left( \lambda -i\omega \right) ^{2} \\ 
\omega _{0}^{2} \\ 
2\omega _{0}\left( \lambda -i\omega \right)%
\end{array}%
\right) ,  \notag
\end{eqnarray}%
\newline
where $C_{0}$ and $C_{\pm }$ are constants. The corresponding solution of
the initial value problem is given by%
\begin{eqnarray}
\left( 
\begin{array}{c}
\left\langle p^{2}\right\rangle \\ 
\left\langle x^{2}\right\rangle \\ 
\left\langle px+xp\right\rangle%
\end{array}%
\right) &=&\frac{1}{2\omega ^{2}}\left( \omega _{0}\left( \left\langle
p^{2}\right\rangle _{0}+\left\langle x^{2}\right\rangle _{0}\right) -\lambda
\left\langle px+xp\right\rangle _{0}\right) e^{\lambda t}\left( 
\begin{array}{c}
\omega _{0} \\ 
\omega _{0} \\ 
2\lambda%
\end{array}%
\right)  \label{ex12d} \\
&&+\frac{1}{2\omega ^{2}}\left( \frac{\lambda }{\omega _{0}}\left\langle
px+xp\right\rangle _{0}+\frac{\omega ^{2}-\lambda ^{2}}{\omega _{0}^{2}}%
\left\langle x^{2}\right\rangle _{0}-\left\langle p^{2}\right\rangle
_{0}\right)  \notag \\
&&\quad \times e^{\lambda t}\left( 
\begin{array}{c}
\left( \lambda ^{2}-\omega ^{2}\right) \cos 2\omega t-2\lambda \omega \sin
2\omega t \\ 
\omega _{0}^{2}\cos 2\omega t \\ 
2\lambda \omega _{0}\cos 2\omega t-2\omega _{0}\omega \sin 2\omega t%
\end{array}%
\right)  \notag \\
&&+\frac{1}{2\omega _{0}\omega }\left( \left\langle px+xp\right\rangle _{0}-%
\frac{2\lambda }{\omega _{0}}\left\langle x^{2}\right\rangle _{0}\right) 
\notag \\
&&\quad \times e^{\lambda t}\left( 
\begin{array}{c}
2\lambda \omega \cos 2\omega t+\left( \lambda ^{2}-\omega ^{2}\right) \sin
2\omega t \\ 
\omega _{0}^{2}\sin 2\omega t \\ 
2\omega _{0}\omega \cos 2\omega t+2\lambda \omega _{0}\sin 2\omega t%
\end{array}%
\right) .  \notag
\end{eqnarray}

The mechanical energy operator $E$ can be conveniently introduced as the
Hamiltonian of our shifted linear harmonic oscillator (\ref{g7}):%
\begin{equation}
E=H_{0}=\frac{\omega _{0}}{2}\left( p^{2}+x^{2}\right) -\frac{\lambda }{2}%
\left( px+xp\right) ,  \label{ex8}
\end{equation}%
so that%
\begin{equation}
H=H_{0}+i\frac{\lambda }{2}.  \label{ex9}
\end{equation}%
Then%
\begin{eqnarray}
\frac{d}{dt}\left\langle E\right\rangle &=&\frac{\omega _{0}}{2}\left( \frac{%
d}{dt}\left\langle p^{2}\right\rangle +\frac{d}{dt}\left\langle
x^{2}\right\rangle \right)  \label{ex10} \\
&&-\frac{\lambda }{2}\frac{d}{dt}\left\langle px+xp\right\rangle  \notag \\
&=&\lambda \left\langle \frac{\omega _{0}}{2}\left( p^{2}+x^{2}\right) -%
\frac{\lambda }{2}\left( px+xp\right) \right\rangle  \notag
\end{eqnarray}%
with the help of our system (\ref{ex12}). Therefore,%
\begin{equation}
\frac{d}{dt}\left\langle E\right\rangle =\lambda \left\langle E\right\rangle
,\qquad \left\langle E\right\rangle =\left\langle E\right\rangle
_{0}e^{\lambda t}  \label{ex11}
\end{equation}%
for the expectation value of the mechanical energy of the damped oscillator
under consideration.\medskip

The case of the second Hamiltonian:%
\begin{equation}
H=\frac{\omega _{0}}{2}\left( p^{2}+x^{2}\right) -\lambda xp=H_{0}-i\frac{%
\lambda }{2},  \label{ex12e}
\end{equation}%
which is the Hermitian adjoint of the Hamiltonian (\ref{ex4}), is similar.
Here%
\begin{equation*}
H^{n+1}-H^{\dagger }H^{n}=\left( H-H^{\dagger }\right) H^{n}=\lambda \left[
p,x\right] H^{n}=-i\lambda H^{n}
\end{equation*}%
and%
\begin{equation}
\frac{d}{dt}\left\langle H^{n}\right\rangle =-\lambda \left\langle
H^{n}\right\rangle ,\qquad \left\langle H^{n}\right\rangle =\left\langle
H^{n}\right\rangle _{0}e^{-\lambda t}\qquad \left( n=0,1,2,...\right) .
\label{ex12ea}
\end{equation}%
Moreover,%
\begin{eqnarray}
p^{2}H-H^{\dagger }p^{2} &=&\frac{\omega _{0}}{2}\left[ p^{2},x^{2}\right]
+\lambda p\left[ x,p\right] p  \label{ex12aa} \\
&=&i\lambda p^{2}-i\omega _{0}\left( px+xp\right) ,  \notag
\end{eqnarray}%
\begin{eqnarray}
x^{2}H-H^{\dagger }x^{2} &=&\frac{\omega _{0}}{2}\left[ x^{2},p^{2}\right]
+\lambda \left[ p,x^{3}\right]  \label{ex12ab} \\
&=&-3i\lambda x^{2}+i\omega _{0}\left( px+xp\right) ,  \notag
\end{eqnarray}%
\begin{eqnarray}
&&\left( px+xp\right) H-H^{\dagger }\left( px+xp\right)  \label{ex12bb} \\
&&\quad =\frac{\omega _{0}}{2}\left( \left[ p,x^{3}\right] +\left[ x,p^{3}%
\right] \right)  \notag \\
&&\qquad +\frac{\omega _{0}}{2}\left( p\left[ x,p\right] p-x\left[ x,p\right]
x\right)  \notag \\
&&\qquad \quad -\lambda \left( \left( xp\right) ^{2}-\left( px\right)
^{2}\right)  \notag \\
&&\quad =2i\omega _{0}\left( p^{2}-x^{2}\right) -i\lambda \left(
px+xp\right) ,  \notag
\end{eqnarray}%
and the corresponding system has the form%
\begin{eqnarray}
&&\frac{d}{dt}\left\langle p^{2}\right\rangle =\lambda \left\langle
p^{2}\right\rangle -\omega _{0}\left\langle px+xp\right\rangle ,  \notag \\
&&\frac{d}{dt}\left\langle x^{2}\right\rangle =-3\lambda \left\langle
x^{2}\right\rangle +\omega _{0}\left\langle px+xp\right\rangle ,
\label{ex12ba} \\
&&\frac{d}{dt}\left\langle px+xp\right\rangle =2\omega _{0}\left(
\left\langle p^{2}\right\rangle -\left\langle x^{2}\right\rangle \right)
-\lambda \left\langle px+xp\right\rangle .  \notag
\end{eqnarray}%
The change $p\leftrightarrow x,$ $\lambda \rightarrow -\lambda ,$ $\omega
_{0}\rightarrow -\omega _{0}$ transforms formally this system back into (\ref%
{ex12}). This observation allows us to obtain solution of the initial value
problem from the previous solution given by (\ref{ex12d}). For the
mechanical energy operator $E$ introduced by equation (\ref{ex8}) one gets%
\begin{equation}
\frac{d}{dt}\left\langle E\right\rangle =-\lambda \left\langle
E\right\rangle ,\qquad \left\langle E\right\rangle =\left\langle
E\right\rangle _{0}e^{-\lambda t}  \label{ex12bab}
\end{equation}%
with the help of (\ref{ex12ba}).\medskip

The case of a general variable quadratic Hamiltonian of the form%
\begin{equation}
H=a\left( t\right) p^{2}+b\left( t\right) x^{2}+c\left( t\right) px+d\left(
t\right) xp,  \label{gen1}
\end{equation}%
where $a\left( t\right) ,$ $b\left( t\right) ,$ $c\left( t\right) ,$ $%
d\left( t\right) $ are real-valued functions of time only, is considered in
a similar fashion. One gets%
\begin{equation}
H^{n+1}-H^{\dagger }H^{n}=\left( H-H^{\dagger }\right) H^{n}=\left(
c-d\right) \left[ p,x\right] H^{n}=i\left( d-c\right) H^{n}  \label{gen2}
\end{equation}%
and%
\begin{equation}
\frac{d}{dt}\left\langle H^{n}\right\rangle =\left\langle \frac{\partial
H^{n}}{\partial t}\right\rangle +\left( d\left( t\right) -c\left( t\right)
\right) \left\langle H^{n}\right\rangle .  \label{gen4}
\end{equation}%
The cases $n=0$ and $n=1$ result in%
\begin{equation}
\left\langle 1\right\rangle =\left\langle 1\right\rangle _{0}\exp \left(
\int_{0}^{t}\left( d\left( \tau \right) -c\left( \tau \right) \right) \
d\tau \right) \label{gen4a}
\end{equation}%
and%
\begin{equation}
\frac{d}{dt}\left\langle H\right\rangle =\left\langle \frac{\partial H}{%
\partial t}\right\rangle +\left( d\left( t\right) -c\left( t\right) \right)
\left\langle H\right\rangle ,\label{gen4b}
\end{equation}%
respectively.\medskip 

Moreover,%
\begin{eqnarray}
p^{2}H-H^{\dagger }p^{2} &=&b\left[ p^{2},x^{2}\right] +c\left[ p^{3},x%
\right] +dp\left[ p,x\right] p  \label{gen5} \\
&=&-i\left( 3c+d\right) p^{2}-2ib\left( px+xp\right) ,  \notag
\end{eqnarray}%
\begin{eqnarray}
x^{2}H-H^{\dagger }x^{2} &=&a\left[ x^{2},p^{2}\right] +cx\left[ x,p\right]
x+d\left[ x^{3},p\right]   \label{gen6} \\
&=&i\left( 3d+c\right) x^{2}+2ia\left( px+xp\right) ,  \notag
\end{eqnarray}%
\begin{eqnarray}
&&\left( px+xp\right) H-H^{\dagger }\left( px+xp\right)   \label{gen7} \\
&&\quad =a\left( \left[ x,p^{3}\right] +p\left[ x,p\right] p\right)   \notag
\\
&&\qquad +b\left( \left[ p,x^{3}\right] +x\left[ p,x\right] x\right)   \notag
\\
&&\qquad \quad +\left( c-d\right) \left( \left( px\right) ^{2}-\left(
xp\right) ^{2}\right)   \notag \\
&&\quad =4iap^{2}-4ibx^{2}-i\left( c-d\right) \left( px+xp\right) ,  \notag
\end{eqnarray}%
and the corresponding system has the form%
\begin{eqnarray}
&&\frac{d}{dt}\left\langle p^{2}\right\rangle =-\left( 3c+d\right)
\left\langle p^{2}\right\rangle -2b\left\langle px+xp\right\rangle ,  \notag
\\
&&\frac{d}{dt}\left\langle x^{2}\right\rangle =\left( c+3d\right)
\left\langle x^{2}\right\rangle +2a\left\langle px+xp\right\rangle ,
\label{gen8} \\
&&\frac{d}{dt}\left\langle px+xp\right\rangle =4a\left\langle
p^{2}\right\rangle -4b\left\langle x^{2}\right\rangle +\left( d-c\right)
\left\langle px+xp\right\rangle .  \notag
\end{eqnarray}%
We have used the familiar identities%
\begin{equation}
\left[ x,p\right] =i,\qquad \left( xp\right) ^{2}-\left( px\right)
^{2}=i\left( px+xp\right) ,  \label{gen9}
\end{equation}%
\begin{equation}
\left[ x^{2},p^{2}\right] =2i\left( px+xp\right) ,\qquad \left[ x,p^{3}%
\right] =3ip^{2},\qquad \left[ x^{3},p\right] =3ix^{2}  \label{gen10}
\end{equation}%
once again.

\section{A Relation with the Classical Damped Oscillations}

Application of formula (\ref{ex1}) to the position $x$ and momentum $p$
operators allows to modify the Ehrenfest theorem \cite{Ehrenfest}, \cite%
{Merz}, \cite{Schiff} for the models of damped oscillators under
consideration. For the Hamiltonian (\ref{ex4}) one gets%
\begin{eqnarray}
xH-H^{\dagger }x &=&\frac{\omega _{0}}{2}\left[ x,p^{2}\right] =i\omega
_{0}p,  \label{cl1} \\
pH-H^{\dagger }p &=&\frac{\omega _{0}}{2}\left[ p,x^{2}\right] +\lambda %
\left[ x,p^{2}\right] =-i\omega _{0}x+2i\lambda p  \label{cl2}
\end{eqnarray}%
and%
\begin{equation}
\frac{d}{dt}\left\langle x\right\rangle =\omega _{0}\left\langle
p\right\rangle ,\qquad \frac{d}{dt}\left\langle p\right\rangle =-\omega
_{0}\left\langle x\right\rangle +2\lambda \left\langle p\right\rangle .
\label{cl3}
\end{equation}%
Elimination of the expectation value $\left\langle p\right\rangle $ from
this system results in%
\begin{equation}
\frac{d^{2}}{dt^{2}}\left\langle x\right\rangle -2\lambda \frac{d}{dt}%
\left\langle x\right\rangle +\omega _{0}^{2}\left\langle x\right\rangle =0,
\label{cl4}
\end{equation}%
which is a classical equation of motion for a damped oscillator \cite%
{BatemanPDE}, \cite{Lan:Lif}.\medskip

For the second Hamiltonian (\ref{ex12e}) we obtain%
\begin{equation}
\frac{d}{dt}\left\langle x\right\rangle =\omega _{0}\left\langle
p\right\rangle -2\lambda \left\langle x\right\rangle ,\qquad \frac{d}{dt}%
\left\langle p\right\rangle =-\omega _{0}\left\langle x\right\rangle ,
\label{cl5}
\end{equation}%
which gives%
\begin{equation}
\frac{d^{2}}{dt^{2}}\left\langle x\right\rangle +2\lambda \frac{d}{dt}%
\left\langle x\right\rangle +\omega _{0}^{2}\left\langle x\right\rangle =0
\label{cl6}
\end{equation}
in a similar fashion.

Finally, our model of the shifted harmonic oscillator (\ref{g8}), when the
Hamiltonian is given by (\ref{ex8}), results in%
\begin{equation}
\frac{d^{2}}{dt^{2}}\left\langle x\right\rangle +\left( \omega
_{0}^{2}-\lambda ^{2}\right) \left\langle x\right\rangle =0.  \label{c7}
\end{equation}%
We leave the details to the reader.

\section{The Third Model}

For the time-dependent Schr\"{o}dinger equation with variable quadratic
Hamiltonian:%
\begin{equation}
i\frac{\partial \psi }{\partial t}=\frac{\omega _{0}}{2}\left( -e^{-2\lambda
t}\frac{\partial ^{2}\psi }{\partial x^{2}}+e^{2\lambda t}x^{2}\psi \right) ,
\label{sm1}
\end{equation}%
where $a=\left( \omega _{0}/2\right) e^{-2\lambda t},$ $b=\left( \omega
_{0}/2\right) e^{2\lambda t}$ and $c=d=0,$ the characteristic equation takes
the form (\ref{fm2}) with the same solution (\ref{fm3}). The corresponding
propagator has the form (\ref{in2}) with%
\begin{eqnarray}
\alpha \left( t\right) &=&\frac{\omega \cos \omega t-\lambda \sin \omega t}{%
2\omega _{0}\sin \omega t}e^{2\lambda t},  \label{sm2} \\
\beta \left( t\right) &=&-\frac{\omega }{\omega _{0}\sin \omega t}e^{\lambda
t},  \label{sm3} \\
\gamma \left( t\right) &=&\frac{\omega \cos \omega t+\lambda \sin \omega t}{%
2\omega _{0}\sin \omega t}.  \label{sm4}
\end{eqnarray}%
This can be derived directly from equations (\ref{in2})--(\ref{in8}) with
the help of identity (\ref{fm7a}). We leave the details to the reader. It is
worth noting that equation (\ref{sm1}) can be obtain by introducing a
variable unit of length $x\rightarrow xe^{\lambda t}$ in the Hamiltonian of
the linear oscillator.

\section{Momentum Representation}

The time-dependent Schr\"{o}dinger equations for the damped oscillators are
also solved in the momentum representation. One can easily verify that under
the Fourier transform our first Hamiltonian (\ref{ex4}) takes the form of
the second Hamiltonian (\ref{ex12e}) with $\lambda \rightarrow -\lambda $
and visa versa (see, for example, Ref.~\cite{Cor-Sot:Sus} for more details).
Moreover, the inverses of the corresponding time evolution operators are
obtained by the time reversal. Therefore, all identities of the commutative
evolution diagram introduced in Ref.~\cite{Cor-Sot:Sus} for the modified
oscillators are also valid for the quantum damped oscillators under
consideration. We leave further details to the reader.\medskip

\noindent \textbf{Acknowledgment.\/} We thank Professor Carlos Castillo-Ch%
\'{a}vez for support, valuable discussions and encouragement. One of the
authors (RCS) is supported by the following National Science Foundation
programs: Louis Stokes Alliances for Minority Participation (LSAMP): NSF
Cooperative Agreement No. HRD-0602425 (WAESO LSAMP Phase IV); Alliances for
Graduate Education and the Professoriate (AGEP): NSF Cooperative Agreement
No. HRD-0450137 (MGE@MSA AGEP Phase II).

\end{document}